\documentstyle[12pt,epsfig]{article}

\newcommand{\be}{\begin{equation}}
\newcommand{\ee}{\end{equation}}

 1
\font\elevenrm=cmr10 scaled\magstep 1
 1

\def\bef{\hang\noindent}
\def\lnf{{\bf  Up~Through }}
\def\le{{\bf  In~Up }}
\def\boa{{\bf  Up~Stop }}
\def\bob{{\bf  In~Down }}
\def\bo{{\bf  Up~Stop~+~In~Down }}
\def\atti{{\it attico }}

\begin{document}
\hspace{10cm}
{\bf Vulcano Workshop 1998}

\hspace{9.7cm}
{\bf May 25-30, Vulcano, Italy}

\vspace*{1.2cm}
  \centerline{\bf MEASUREMENT OF THE ATMOSPHERIC NEUTRINO}
  \centerline{\bf INDUCED MUON FLUX WITH THE MACRO DETECTOR}
\vspace{1cm}
\centerline{PAOLO BERNARDINI for the MACRO Collaboration
\footnote{\ The complete collaboration list is given in the paper 
by G. Battistoni in these Proceedings.}}
\vspace{1.4cm}
  \centerline{DIPARTIMENTO di FISICA dell'UNIVERSIT\`A and INFN}
  \centerline{\elevenrm via per Arnesano, 73100 Lecce, Italy}
\vspace{3cm}

\begin{abstract}
The flux of neutrino-induced muons has been measured with
the MACRO detector. Different event topologies have been detected,
due to neutrino interactions in the apparatus and in the rock
below it. The upward-throughgoing muon sample is the larger one and
is generated by neutrinos with an average energy of $\sim 100 \ GeV$. 
The observed upward-throughgoing muons are $26 \ \%$ fewer than
expected and the zenith angle distribution does not fit with the
expected one. Assuming neutrino oscillations, both measurements
suggest maximum mixing and $\Delta m^2$ of a few times $10^{-3} \ eV^2$. 
The other event categories due to interactions of low-energy 
neutrinos (average energy $\sim 4 \ GeV$) have been recently studied 
and the results of these new analyses are presented for the first 
time at this workshop. These data show a regular deficit of observed 
events in each angular bin, as expected assuming neutrino 
oscillations with maximum mixing, in agreement with the analysis of 
the upward-throughgoing muon sample. 
\end{abstract}
\vspace{2.0cm}

\section{Introduction}
The interest in precise measurements of the flux of neutrinos 
produced in cosmic ray cascades in the 
atmosphere has been growing over the last years due to the anomaly 
in the ratio of contained muon neutrino to electron neutrino 
interactions. The observations of Kamiokande (Hirata et al., 1992), 
IMB (Casper et al., 1991; Becker-Szendy et al., 1992) and Soudan~2 
(Allison et al., 1997) are now confirmed by those of SuperKamiokande 
(Fukuda et al., 1998) and the anomaly finds explanation in the scenario 
of neutrino oscillation. 

The effects of neutrino oscillation have to appear also in the higher
energy ranges, as reported by MACRO. The flux of muon neutrinos in 
the energy region
from a few $GeV$ up to hundreds of $GeV$ can be inferred from
measurements of upgoing muons (Ahlen et al., 1995, Ambrosio et al., 
1998, 2nd reference). As a consequence of oscillation,
the flux of upgoing muons should be affected both in the absolute
number of events and in the shape of the zenith angle
distribution, with relatively fewer observed events near the
vertical than near the horizontal due to the longer pathlength of
neutrinos from production to observation near the zenith. 

Furthermore the flux of atmospheric muon neutrinos in the region
of a few $GeV$ can be studied looking at muons produced inside the
detector and muons externally produced and stopping inside it. If
the atmospheric neutrino anomalies are the result of neutrino
oscillations, it is expected a reduction in the flux of
upward-going low-energy atmospheric neutrinos of about a factor of
two, but without any distortion in the shape of the angular
distribution. 

Here the measurement about high energy muon neutrino flux is 
presented, together with first results on low-energy neutrino
events in MACRO. 

\section{The MACRO detector}

The MACRO detector (Ahlen et al., 1993) is located in the Gran Sasso 
Laboratory, with a minimum rock overburden of $2700 \ hg/cm^2$. It is 
a large rectangular box ($76.6 \times 12 \times 9.3 \ m^3$) divided 
longitudinally in $6$ supermodules and vertically in a lower and an 
upper part, called {\it attico}. The active elements (see 
Fig.~\ref{fig:topo}) 
are liquid scintillator counters for time measurement and streamer 
tubes for tracking, with $27^\circ$ stereo strip readouts. The lower 
half of the detector is filled with trays of crushed rock absorber 
alternating with streamer tube planes, while the {\it attico} is 
hollow and contains the electronics racks and work areas. The intrinsic 
angular resolution for muons typically ranges from $0.2^\circ$ to 
$1^\circ$ depending on the track length. 
This resolution is lower than the angular spread due to multiple
scattering of muons in the rock. The scintillator system consists
of horizontal and vertical layers. Time and position resolution
for muons in a scintillator box are about $0.5 \ ns$ and $11 \ cm$, 
respectively. 

Thanks to its large area, fine tracking granularity and electronics 
symmetry with respect to upgoing and downgoing flight direction, the 
MACRO detector is a proper tool for the study of upward-travelling 
muons, generated by external interactions. Its mass permits also
to collect a statistically significant number of neutrino events
due to internal interactions. 

\section{Neutrino events in MACRO}

\begin{figure}[thb]
 \begin{center}
  \vspace{2cm}
  \hspace{-2cm}
  \mbox{\epsfig{file=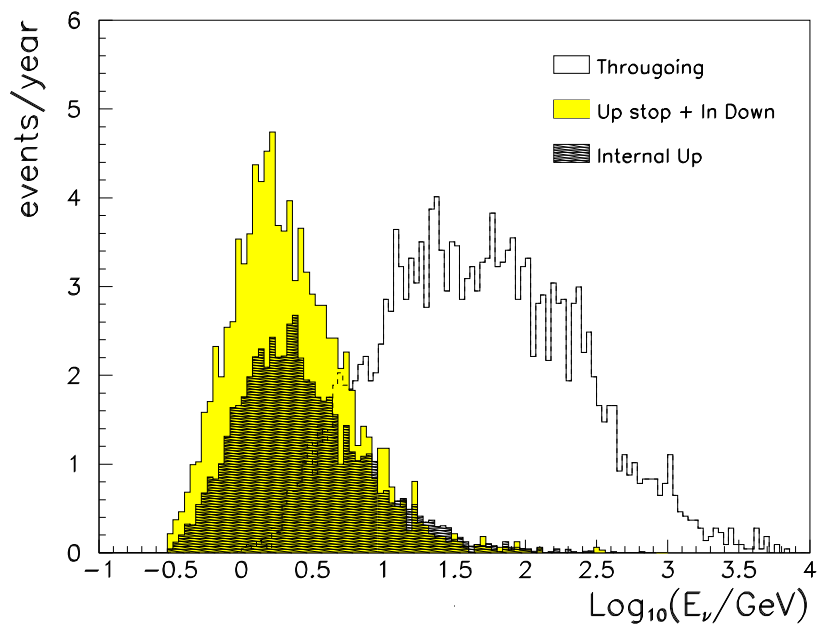,width=17cm}}
  \vspace{-4.5cm}
  \caption{\em {Distributions of the parent neutrino energy giving 
  rise to different kinds of events, estimated by means of Monte 
  Carlo simulation using the same cuts applied to the data. The 
  average energy is $\sim 100 \ GeV$ for \lnf sample and 
  $\sim 4 \ GeV$ for \le and \bo samples. 
  \label{fig:entopo}}}
 \end{center}
\end{figure}

%
Fig.~\ref{fig:topo} displays the different kinds of neutrino events here 
analyzed. Most of the detected particles are muons generated in $\nu_\mu$ 
Charged Current interactions. Figure~\ref{fig:entopo} shows the 
parent neutrino energy distribution for the different event topologies :
\begin{figure}[t]
 \begin{center}
  \vspace{-0.65cm}
  \mbox{\epsfig{file=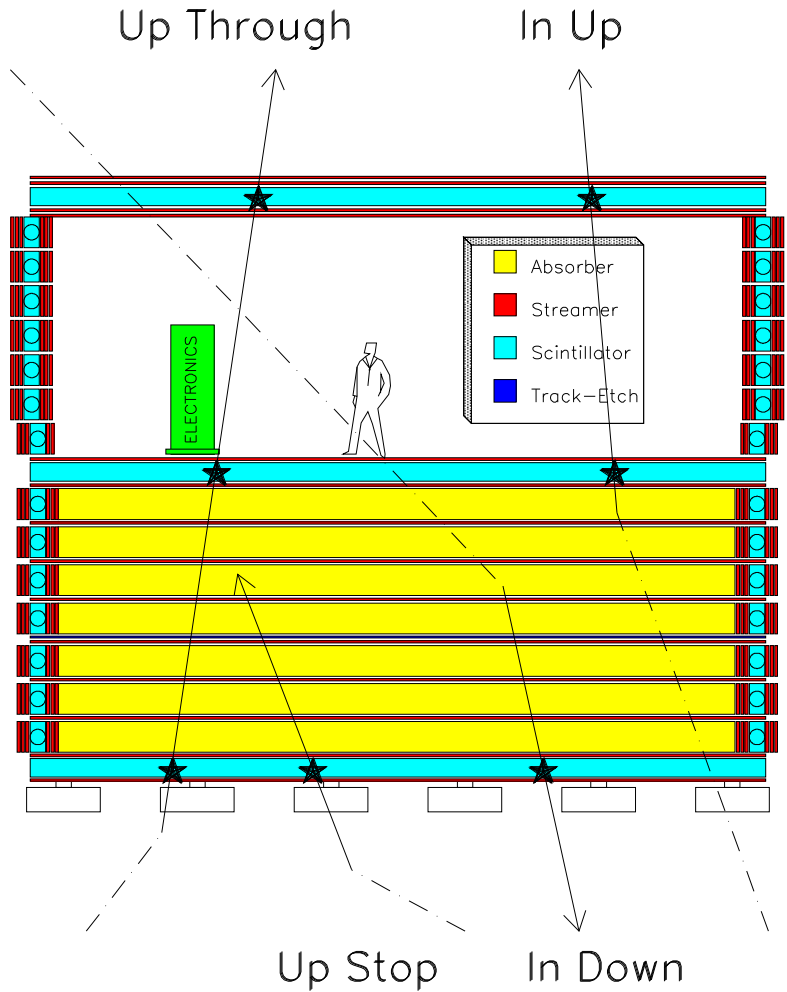,width=15cm}}
  \vspace{-1.0cm}
  \caption{\em {Cross view of the detector and topology of the 
  neutrino induced events. The stars, the dashed lines and the full
  lines indicate scintillator hits, neutrino paths and charged 
  particle paths, respectively. \label{fig:topo}}}
 \end{center}
\end{figure}
\begin{enumerate}
    \item \lnf - These tracks come from interactions in the rock 
        below MACRO and cross the whole detector ($E_\mu > 1 \ GeV$). 
        The time information provided by scintillator counters 
        permits to know the flight direction (time-of-flight method).
        The data have been collected in three periods, with different 
        detector configurations. In the first two periods (March 
        1989 -- November 1991, December 1992 -- June 1993) only lower 
        parts of MACRO were working (Ahlen et al., 1995). In the last 
        period (April 1994 -- November 1997) also the \atti was in 
        acquisition.
    \item \le - These partially contained events come from $\nu$ 
        interactions inside the apparatus. Also in this case the 
        time-of-flight method is applied to identify the events,
        thanks to the \atti scintillator layers. Hence only the data 
        collected with the \atti (live-time $\sim 3$ years) have
        been used in this analysis. About $13 \ \%$ of events are 
        estimated to be induced by Neutral Currents or $\nu_e$ 
        CC interactions.
    \item \bo - This sample is composed by two subsamples :
        external interactions with upward-going track
        stopping in the detector ({\bf Up Stop}), neutrino-induced 
        downgoing tracks with vertex in lower part of MACRO ({\bf In 
        Down}).
        These events are identified by means of topological criteria.
        The lack of time information prevents to distinguish the two 
        subsamples. Anyway an almost equal number of \boa and \bob is
        expected if neutrinos do not oscillate. Neutral Currents and 
        $\nu_e$ CC interactions constitute $\sim 10 \ \%$ of the 
        sample. The analyzed data have been collected with the whole 
        detector with an effective live-time of $\sim 3$ years.
\end{enumerate}


\section{Analysis procedure and results}

\begin{figure}[thb]
 \begin{center}
  \vspace{-0.5cm}
  \mbox{\epsfig{file=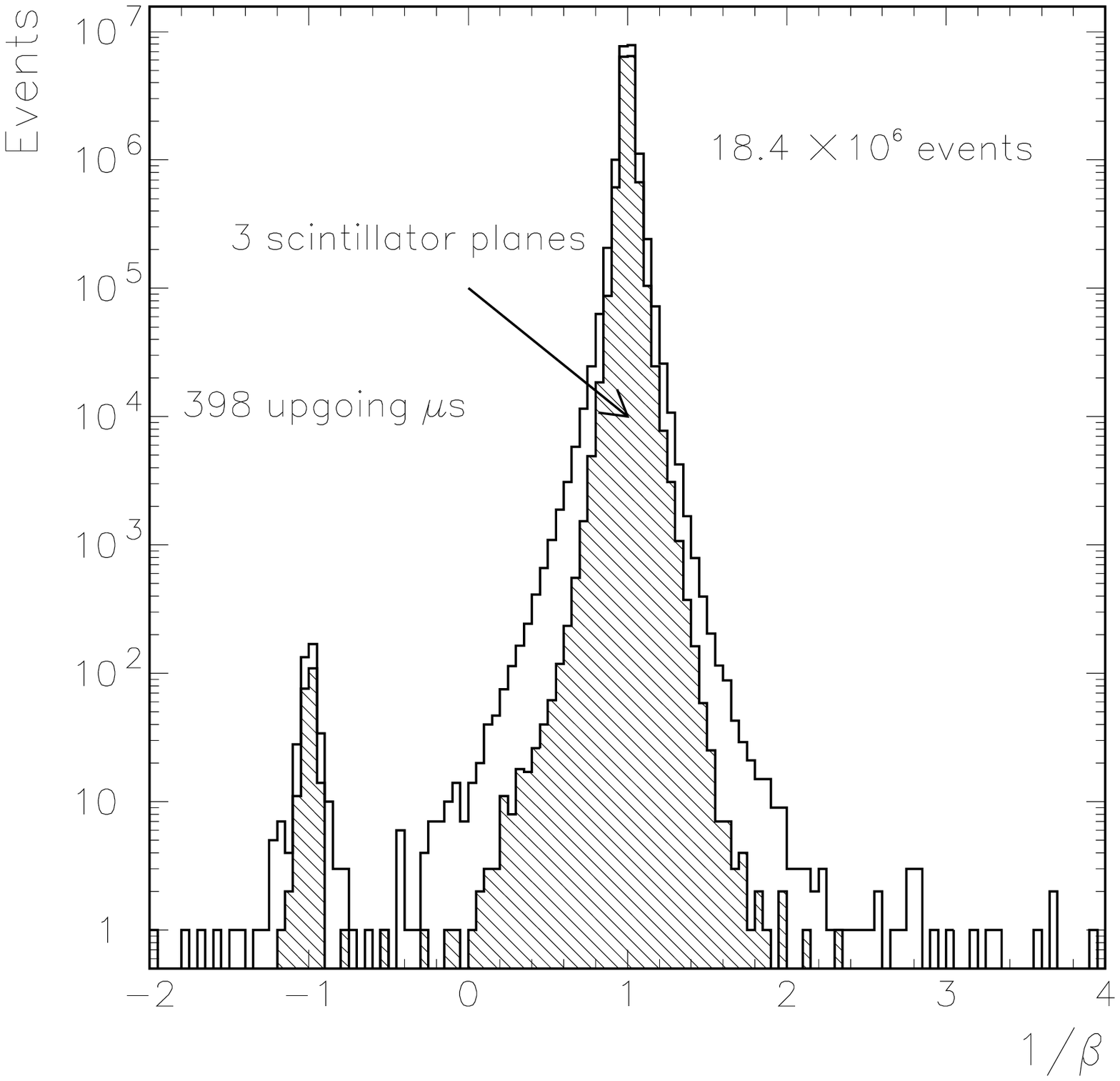,width=7.6cm}}
  \mbox{\epsfig{file=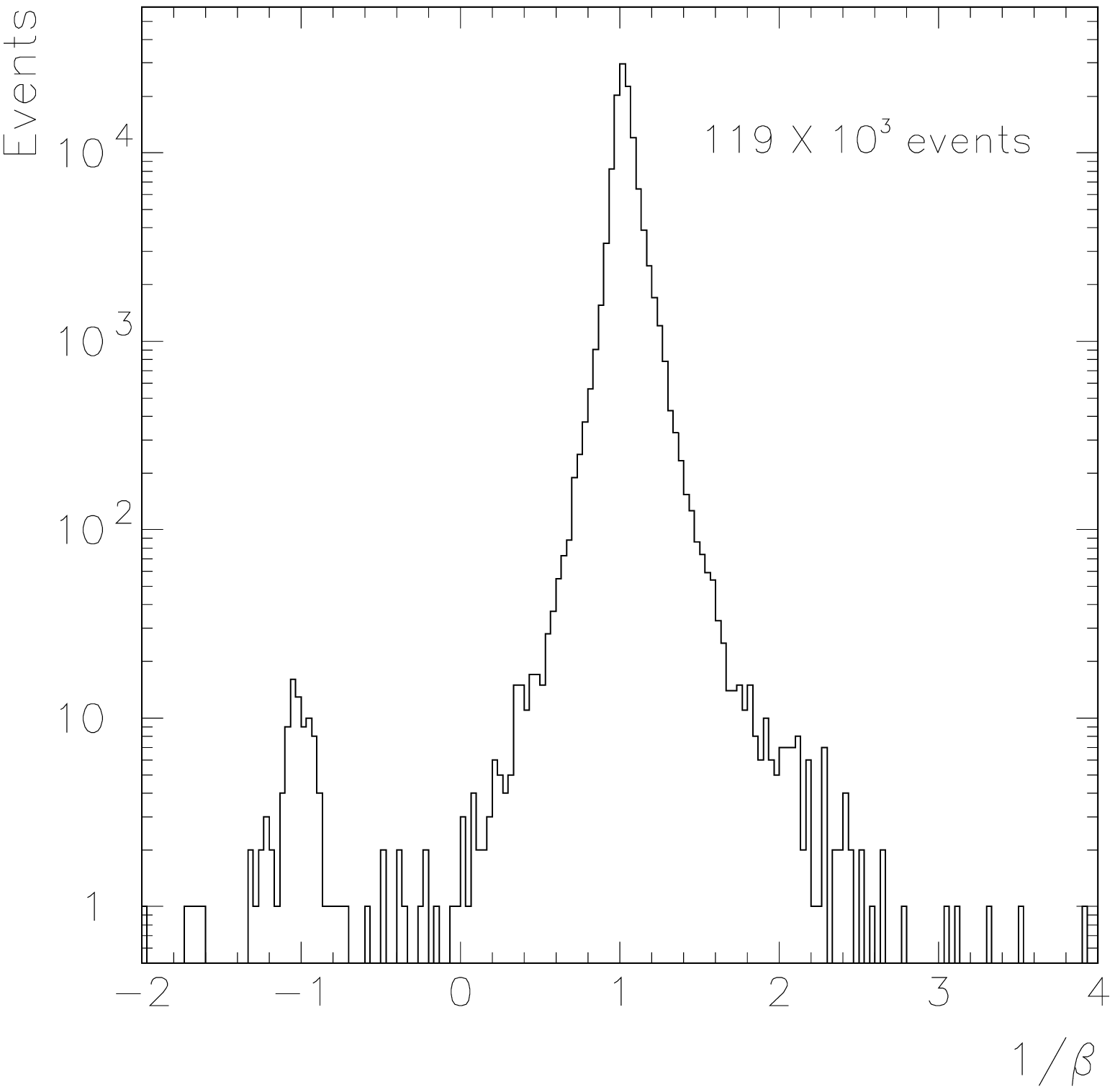,width=7.6cm}}
  \caption{\em {$1/\beta$ distributions after all analysis cuts.
  In the first plot the \lnf data collected in the third period
  (lower MACRO + \atti) are shown, the shaded part indicates events 
  whose $1/\beta$ value is calculated by means of a fit of three 
  time measurements on different scintillator planes. In the second 
  plot the whole \le sample is shown. 
  \label{fig:sbeta}}}
 \end{center}
\end{figure}

The time-of-flight method uses the formula
\be \label{beta}
\frac{1}{\beta} = \frac{c \times (T_1 - T_2)}{L}, \ee
where $T_1$ and $T_2$ are the times measured in lower and higher 
scintillator planes, respectively, and $L$ is the path between the 
two scintillators. Therefore $1/\beta$ results roughly $+1$ for 
downgoing tracks and $-1$ for upgoing tracks. Several cuts are 
imposed to remove backgrounds from radioactivity and showering 
events which may cause failure in time reconstruction. Another cut 
is applied to the \lnf sample requiring the crossing of at least 
$200 \ g/cm^2$ of material in the apparatus in order to reduce 
the background due to low-energy charged upgoing particles 
produced at large angles by downgoing muons (Ambrosio et al., 1998).
After all analysis cuts the signal peaks with $1/\beta \sim -1$ 
are well isolated for the first two samples (see Fig.~\ref{fig:sbeta}). 

The \bo events are identified via topological constraints. The main 
requirement is the presence of a reconstructed track crossing the 
bottom scintillator layer. All the track hits must be at least 
$1 \ m$ far from the supermodule walls. The criteria used to verify 
that the event vertex (or $\mu$ stop point) is inside the detector 
are similar to those used for the \le search. The probability 
that an atmospheric muon produces a background event is negligible.
To reject ambiguous and/or wrongly tracked events which survived 
automated analysis cuts, real and simulated events were randomly 
merged and directly scanned with the MACRO Event Display.

Expected rates and angular distributions have been estimated 
assuming the atmospheric $\nu$ flux calculated by the Bartol group
(Gaisser and Stanev, 1995, Agrawal et al., 1996). The estimate of 
$\nu$ cross-section was based on parton set $S_1$ from (Morfin and 
Tung, 1991), taking into account also low-energy effects (Lipari et 
al., 1995) for \le and \bo samples. The propagation of muons through 
the rock was taken from 
(Lohmann et al., 1985). The uncertainty on the expected muon flux 
is estimated $17 \ \%$ for \lnf events and $25 \ \%$ for the other
events. The apparatus and the data acquisition are fully reproduced 
in a GEANT (Brun et al., 1992) based Monte Carlo program and the 
simulated data are processed by means of the same analysis chain used 
for real data. 
Particular care has been taken to minimize the systematic uncertainty 
in the detector acceptance simulation. For the \lnf sample, several 
different analyses and acceptance calculations, including separate 
electronic and data acquisition systems, have been compared. For each 
sample, trigger and streamer tube efficiency, background subtraction, 
effects of analysis cuts have been in detail studied. The systematic 
error on the total number of events due to the acceptance has been 
estimated $6 \ \%$ for \lnf sample. The uncertainty is higher 
($10 \ \%$) for low-energy samples because it depends strongly on data 
taking conditions, analysis algorithm efficiency and mass of the 
detector.

\begin{figure}[thb]
 \begin{center}
  \mbox{\epsfig{file=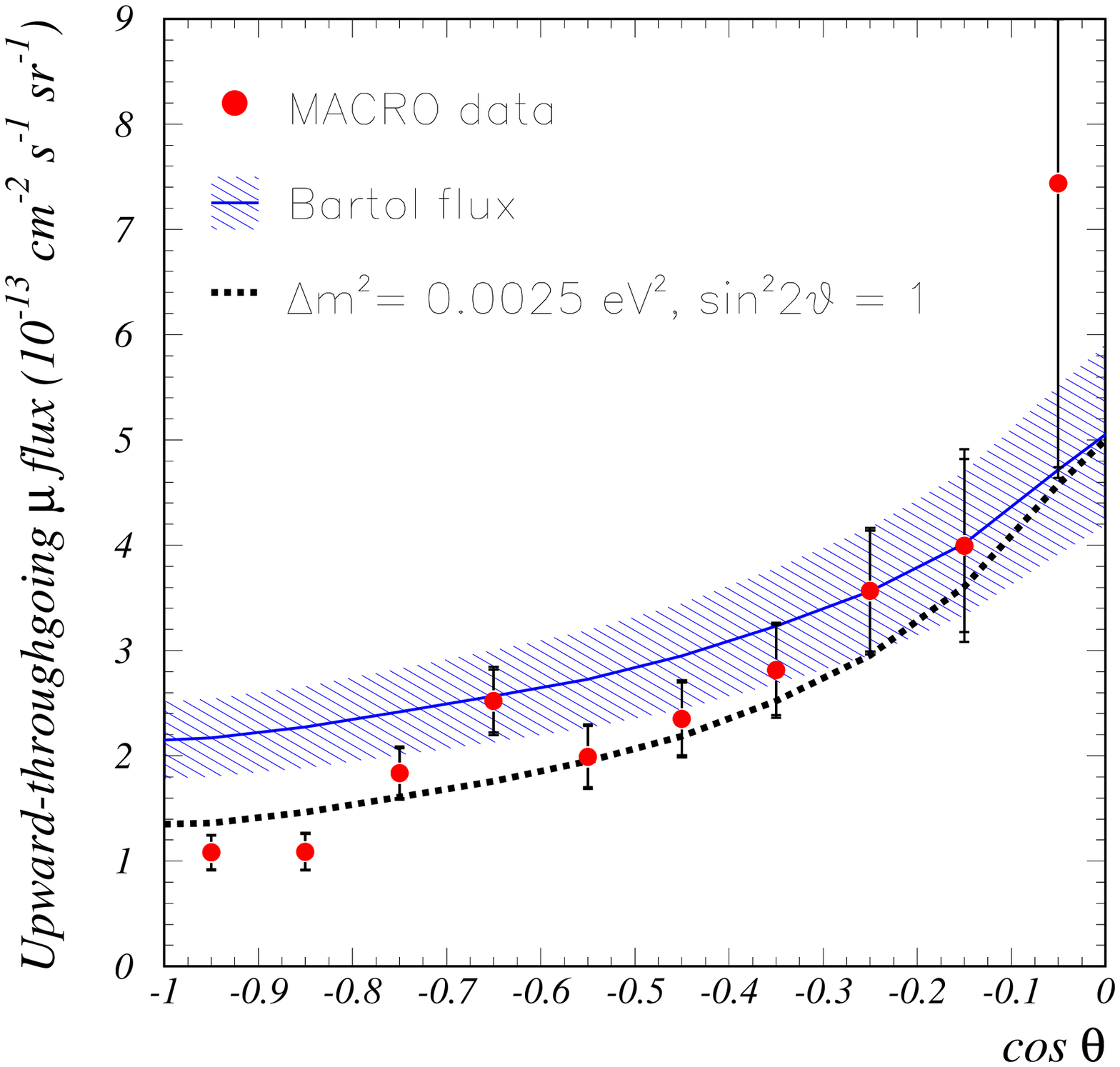,width=7.9cm}}
  \mbox{\epsfig{file=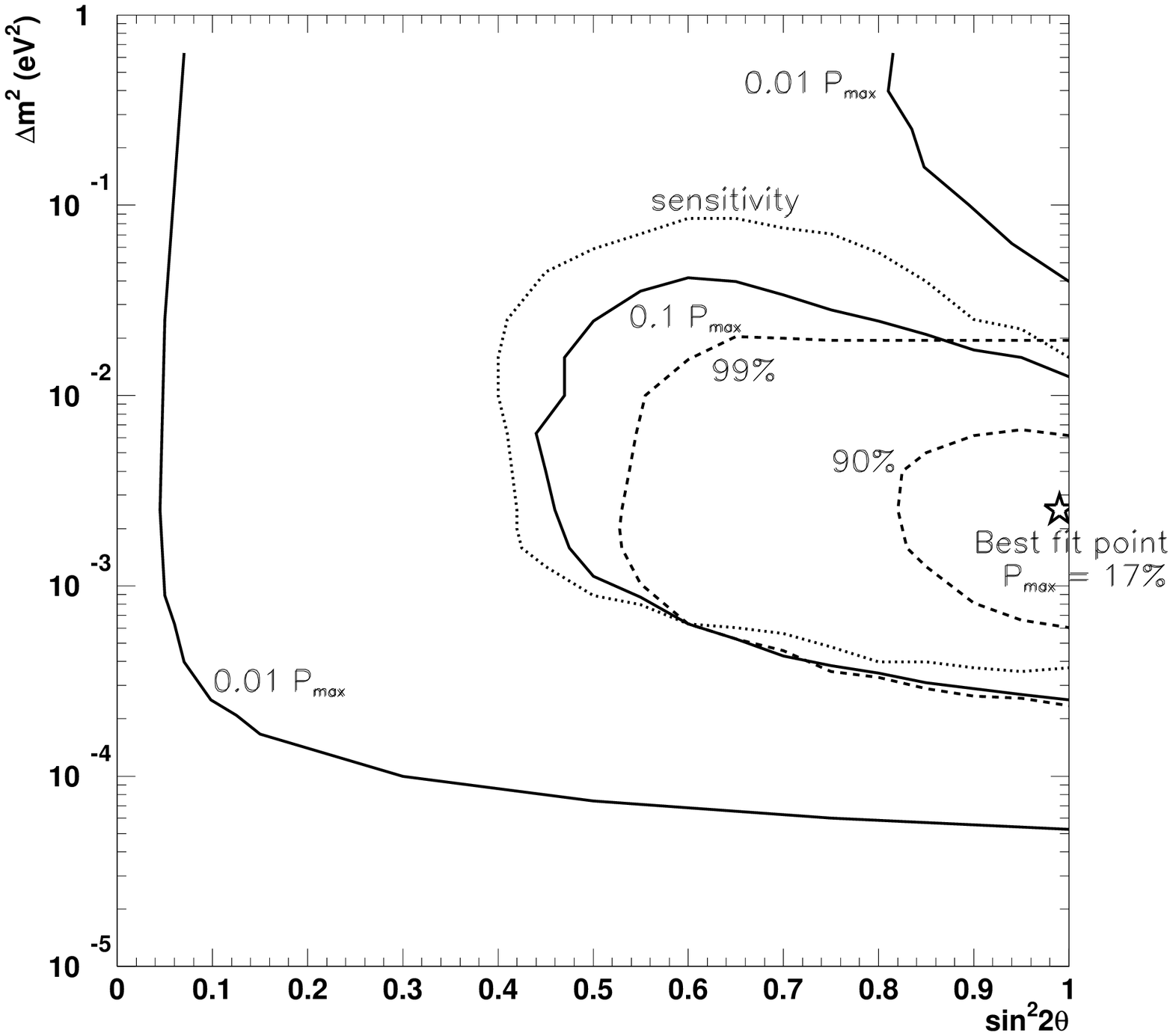,width=7.36cm}}
  \vspace{-0.4cm}
  \caption{\em {In the first plot it is displayed the comparison 
  between measured and expected fluxes, for the \lnf sample with a 
  muon energy threshold of $1 \ GeV$. The solid curve and the shaded 
  region show the expectation for no oscillation and its uncertainty.
  The dashed line shows the predicition assuming $\nu$ oscillation 
  with maximum mixing and $\Delta m^2 = 2.5 \times 10^{-3} \ eV^2$. 
  In the second plot the results of the combined analysis for \lnf 
  sample are shown in terms of oscillation parameters (see the text). 
  \label{fig:flux}}}
 \end{center}
\end{figure}

In the \lnf sample 479 events are in the signal range 
($0.25$ around $1/\beta = -1$). After the subtraction of the estimated 
backgrounds, the observed number of events becomes $451$. For this 
sample $612$ events are expected and the ratio observation/expectation 
is reported in Table~\ref{tab:nosci}. 
Fig.~\ref{fig:flux} shows the zenith angle distribution of the measured 
flux 
compared with the expectation. The error bars on the data show the 
statistical errors with an extension due to the systematic errors, added 
in quadrature. 
The observed zenith distribution does not fit well 
with the expectation, giving a maximum $\chi^2$ probability of only 
$0.1 \ \%$. The observed number of events and the shape of the zenith 
distribution can be explained in the scenario of $\nu_\mu \rightarrow 
\nu_\tau$ oscillation with a best fit point in the unphysical 
range ($\sin^2 2\theta_{mix} > 1$). Both measurements independently 
yield very
close mixing parameter values. The second plot in Fig.~\ref{fig:flux} 
shows the best fit point in the physical region 
($\sin^2 2\theta_{mix} = 1$ and 
$\Delta m^2 = 2.5 \times 10^{-3} \ eV^2$) obtained combining 
event number and angular shape analysis. The probability associated to 
this point is not so high ($17 \ \%$) because the probability of the 
zenith distribution is still low ($5 \ \%$), due to the relatively few 
events in the region $-1 < cos\theta < -0.8$ compared with the number 
of events in $-0.8 < cos\theta$. 
In the second plot of Fig.~\ref{fig:flux} the solid lines show the 
contours for $10 \ \%$ and $1 \ \%$ of the best fit probability. 
The dashed lines show the allowed regions at $90 \ \%$ and $99 \ \%$ c.l. 
evaluated according to the prescription by (Feldman and Cousins, 1998). 
The dotted line shows the sensitivity which is the $90 \ \%$ c.l. 
contour which would result from the preceding prescription if the data 
and the M.C. prediction happened to be in perfect agreement at the 
best fit point.
\begin{table}[b]
 \begin{center}
  \begin{tabular}{|c||c|c|c|c||c|} \hline
Topology & Ratio & Statist. & Syst. & Theor. &    Ratio with     \\
         &       &  error   & error & error  & $\nu$ oscillation \\\hline
  \lnf   & 0.74  &  0.036   & 0.046 & 0.13   &        1.05       \\
  \le    & 0.59  &  0.06    & 0.06  & 0.15   &        1.02       \\
  \bo    & 0.75  &  0.07    & 0.08  & 0.19   &        0.98       \\\hline
  \end{tabular}
  \caption{\em {Ratios of observed on expected number of events
     for different event topologies. In the last column the ratio 
     is calculated assuming neutrino oscillation with the parameters 
     suggested by the \lnf sample ($\sin^2 2\theta_{mix} = 1$, 
     $\Delta m^2 = 2.5 \times 10^{-3} \ eV^2$). \label{tab:nosci}}}
 \end{center} 
\end{table} 

\begin{figure}[ht]
 \begin{center}
  \mbox{\epsfig{file=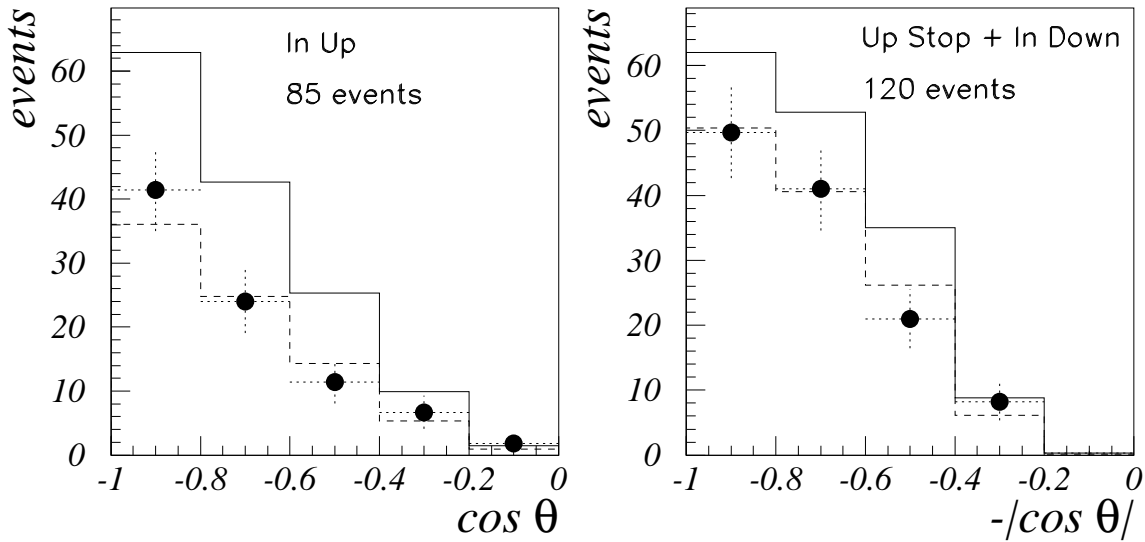,width=17cm}}
  \vspace{-8.7cm}
  \caption{\em {Comparison between measured and expected number of
  low-energy events versus $cos \theta$ (\le in the first plot, \bo
  in the second one). The dashed line is obtained assuming neutrino 
  oscillation with the parameters suggested by \lnf sample. In the
  second plot the absolute value of $cos\theta$ is used because
  the flight direction is unknown.
  \label{fig:doppia}}}
 \end{center}
\end{figure}

The ratios of the observed number of events to the expectation and the 
angular distributions of the \le and \bo data samples are reported 
in Table~\ref{tab:nosci} and in Fig.~\ref{fig:doppia}. The low-energy 
$\nu_\mu$ samples show an uniform deficit of the measured number of 
events over the whole angular distribution with respect to the 
predictions based on the absence of neutrino oscillations. 
We note a good agreement between the results for low-energy 
and \lnf events. Assuming the oscillation parameters suggested by 
higher energy sample, it is expected a $\sim 50 \ \%$ disappearance 
of $\nu_\mu$ in \le and \boa samples because of the neutrino path 
(thousands of kilometres). No flux reduction is instead expected for 
\bob events whose neutrino path is of the order of tens of kilometres. 
The ratios and the angular distributions estimated assuming the $\nu$ 
oscillation are also reported in Table~\ref{tab:nosci} and in 
Fig.~\ref{fig:doppia}. 



\section{Conclusions}


The \lnf sample shows a deficit of the measured number of events 
with respect to the predictions based on the Bartol flux in absence 
of neutrino oscillations. The deficit is higher near the vertical
direction. Hence the previous results of the MACRO experiment
(Ahlen et al., 1995) are confirmed. A new paper about this item
has been submitted for pubblication (Ambrosio et al., 1998, 2nd
reference).

Also the low-energy neutrino events are fewer than expected and
the deficit is quite uniform over the whole angular range. The 
three data samples are in agreement with a model of 
$\nu_\mu \rightarrow \nu_\tau$ oscillation with maximum mixing 
and $\Delta m^2$ of a few times $10^{-3} \ eV^2$. The combined
analysis of the three different data sets is in progress. 
\section {References}

\bef Agrawal, V., Gaisser, T.K., Lipari, P., Stanev, T.: 1996,
Phys.Rev. {\bf D53}, p. 1314.

\bef Ahlen, S., et al. (MACRO Collaboration): 1993, Nucl. 
Instr. and Methods {\bf A324}, p. 337.

\bef Ahlen, S., et al. (MACRO Collaboration): 1995, Phys. 
Lett. {\bf B357}, p. 481.

\bef Allison, W.W.M., et al. (Soudan Collaboration): 1997, 
Phys. Lett. {\bf B391}, p. 491.

\bef Ambrosio, M., et al. (MACRO Collaboration): 1998, Astroparticle 
Physics {\bf 9}, p. 105.

\bef Ambrosio, M., et al. (MACRO Collaboration): 1998, hep-ex/9807005.

\bef Becker-Szendy, R., et al. (IMB Collaboration): 1992, Phys. 
Rev. {\bf D46}, p. 3720.

\bef Brun, R., et al.: 1992, CERN GEANT User's Guide {\bf DD/EE} 84-1.

\bef Casper, D., et al. (IMB Collaboration): 1991, Phys. 
Rev. Lett. {\bf 66}, p. 2561.

\bef Feldman, G.J., Cousins, R.D.: 1998, Phys. Rev. {\bf D57}, p. 3873.

\bef Fukuda, Y., et al. (SuperKamiokande Collaboration): 1998, 
hep-ex/9807003

\bef Gaisser, T.K., Stanev, T.: 1995, Proceedings of the 24th ICRC 
(Rome), vol. {\bf 1}, p. 694.

\bef Hirata, K.S., et al. (Kamiokande Collaboration): 1992, Phys. 
Lett. {\bf B280}, p. 2.

\bef Lipari, P., Lusignoli, M., Sartogo, F.: 1995, Phys. Rev. Lett.
{\bf 74}, p. 4384.

\bef Lohmann, W., Kopp, R., Voss, R.: 1985, CERN-EP 85-03.

\bef Morfin, J. G., Tung, W. K.: 1991, Z. Phys. {\bf C52}, p. 13.

\end{document}